*Review*

# The CMS Magnetic Field Measuring and Monitoring Systems

Vyacheslav Klyukhin [1,2,*], Austin Ball [2], Felix Bergsma [2], Henk Boterenbrood [3], Benoit Curé [2], Domenico Dattola [4], Andrea Gaddi [2], Hubert Gerwig [2], Alain Hervé [5], Richard Loveless [5], Gary Teafoe [6], Daniel Wenman [5], Wolfram Zeuner [2] and Jerry Zimmerman [6]

[1] Skobeltsyn Institute of Nuclear Physics, Lomonosov Moscow State University, RU-119992 Moscow, Russia
[2] European Organization for Nuclear Research (CERN), CH-1211 Geneva 23, Switzerland; austin.ball@cern.ch (A.B.); felix.bergsma@cern.ch (F.B.); Benoit.Cure@cern.ch (B.C.); Andrea.Gaddi@cern.ch (A.G.); Hubert.Gerwig@cern.ch (H.G.); Wolfram.Zeuner@cern.ch (W.Z.)
[3] Nikhef, 1098XG Amsterdam, The Netherlands; boterenbrood@nikhef.nl
[4] Department of Physics, Polytechnic University of Turin, I-10129 Turin, Italy; domenico.dattola@cern.ch
[5] Department of Physics, University of Wisconsin, Madison, WI 53706, USA; Alain.Herve@cern.ch (A.H.); loveless@hep.wisc.edu (R.L.); dwenman@psl.wisc.edu (D.W.)
[6] FNAL, Batavia, IL 60510-0500, USA; teafoe@fnal.gov (G.T.); jerryz@fnal.gov (J.Z.)
* Correspondence: Vyacheslav.Klyukhin@cern.ch

**Abstract:** This review article describes the performance of the magnetic field measuring and monitoring systems for the Compact Muon Solenoid (CMS) detector. To cross-check the magnetic flux distribution obtained with the CMS magnet model, four systems for measuring the magnetic flux density in the detector volume were used. The magnetic induction inside the 6 m diameter superconducting solenoid was measured and is currently monitored by four nuclear magnetic resonance (NMR) probes installed using special tubes at a radius of 2.9148 m outside the barrel hadron calorimeter at ±0.006 m from the coil median *XY*-plane. Two more NRM probes were installed at the faces of the tracking system at *Z*-coordinates of −2.835 and +2.831 m and a radius of 0.651 m from the solenoid axis. The field inside the superconducting solenoid was precisely measured in 2006 in a cylindrical volume of 3.448 m in diameter and 7 m in length using ten three-dimensional (3D) B-sensors based on the Hall effect (Hall probes). These B-sensors were installed on each of the two propeller arms of an automated field-mapping machine. In addition to these measurement systems, a system for monitoring the magnetic field during the CMS detector operation has been developed. Inside the solenoid in the horizontal plane, four 3D B-sensors were installed at the faces of the tracking detector at distances $X = \pm 0.959$ m and *Z*-coordinates of −2.899 and +2.895 m. Twelve 3D B-sensors were installed on the surfaces of the flux-return yoke nose disks. Seventy 3D B-sensors were installed in the air gaps of the CMS magnet yoke in 11 *XY*-planes of the azimuthal sector at 270°. A specially developed flux loop technique was used for the most complex measurements of the magnetic flux density inside the steel blocks of the CMS magnet yoke. The flux loops are installed in 22 sections of the flux-return yoke blocks in grooves of 30 mm wide and 12–13 mm deep and consist of 7–10 turns of 45 wire flat ribbon cable. The areas enclosed by these coils varied from 0.3 to 1.59 m² in the blocks of the barrel wheels and from 0.5 to 1.12 m² in the blocks of the yoke endcap disks. The development of these systems and the results of the magnetic flux density measurements across the CMS magnet are presented and discussed in this review article.

## 1. Introduction

The main difficulty in large magnetic systems having an extensive flux-return yoke is to characterize the magnetic flux distribution in the steel blocks of the yoke. Continuous measurements of the magnetic flux density in the return yoke are not possible; in common practice, software modelling of the magnetic system using special three-dimensional (3D) computer programs is applied [1].

The Compact Muon Solenoid (CMS) detector [2] at the Large Hadron Collider (LHC) [3] has a heterogeneous solenoid magnet where the created magnetic flux penetrates both nonmagnetic and ferromagnetic materials of the experimental setup. The steel yoke of the magnet is used as magnetized layers that wrap the muons, which allows them to be identified and their momenta to be measured in a muon spectrometer. The precise measurements of the charged particle momenta are performed with the homogeneous magnetic field inside the inner tracking volume.

The CMS magnetic field is provided by a wide-aperture superconducting thin solenoid [4] with a diameter of 6 m and a length of 12.5 m, where a central magnetic flux density of 3.81 T is created by an operational direct current of 18.164 kA [5–7].

The CMS detector provides registration of charged particles in the pseudorapidity region $|\eta| < 2.5$, registration of electrons, positrons, and gamma rays in the region $|\eta| < 3$, registration of hadronic jets in the region $|\eta| < 5.2$, and registration of muons in the region $|\eta| < 2.4$ [2]. Here, the pseudorapidity $\eta$ is determined as $\eta = -ln[tan(\theta/2)]$, where $\theta$ is a polar angle in the detector reference frame.

The origin of the CMS coordinate system is at the centre of the superconducting solenoid; the *X* axis lies in the LHC plane and is directed to the centre of the LHC machine; the *Y* axis is directed upward and is perpendicular to the LHC plane; the *Z* axis is the right-hand triplet with the *X* and *Y* axes and is directed along the vector of magnetic induction generated on the axis of the superconducting coil.

The magnetic flux density in the central part of the CMS detector in a cylindrical volume of 3.448 m in diameter and 7 m in length has been measured with a precision of 7 × $10^{-4}$ with a field-mapping machine [8] before the solenoidal volume was filled with physics detectors. The magnetic flux everywhere outside of this measured volume was calculated by a magnetic field 3D model [1] with the program TOSCA (TwO SCAlar potential method) [9] from Cobham CTS Limited, Wimborne, U.K. This model reproduced the magnetic flux density distribution measured with the field-mapping machine inside the CMS coil to within 0.1% [10].

To verify the magnetic flux distribution calculated in the yoke steel blocks, direct measurements of the magnetic flux density in the selected regions of the yoke were performed during the CMS magnet test in 2006 [7] when four fast discharges of the CMS coil (190 s time constant) were triggered manually to test the magnet protection system. These discharges were used to induce voltages with amplitudes of 0.5–4.5 V in 22 flux loops wound around the yoke blocks in special grooves, 30 mm wide and 12–13 mm deep. The loops have 7–10 turns of 45 wire flat ribbon cable and the cross sections of areas enclosed by the flux loops varied from 0.3 to 1.59 $m^2$ on the yoke barrel wheels and from 0.5 to 1.12 $m^2$ on the yoke endcap disks [11]. An integration technique [12] was developed to reconstruct the average initial magnetic flux density in the cross sections of the steel blocks at full magnet excitation.

At that time, no fast discharge of the CMS magnet from its operational current of 18.164 kA, which corresponds to a central magnetic flux density of 3.81 T, had been performed. To measure the magnetic flux density in the steel blocks of the flux-return yoke at the operational current, several standard linear discharges of the CMS magnet with a current rate as low as 1–1.5 A/s were performed later. To provide these measurements, the voltages induced in the flux loops (with amplitudes of 20–250 mV) were measured with six 16-bit data acquisition (DAQ) modules and integrated offline over time.

In addition to these two measurement systems, the magnetic flux density inside the solenoid and in the air gaps of the yoke was and is monitored by the system of six nuclear magnetic resonance (NMR) probes and by a system of 86 3D B-sensors (Hall probes) during the magnet operation.

The review article is organized as follows: Section 2 describes the development of a technique for measuring the magnetic field inside the CMS solenoid, a system for monitoring the magnetic flux density during the CMS detector operation, the development of a flux loop technique of measurements of the magnetic flux density inside the CMS yoke steel blocks, and provides an analysis of eddy current distributions in the CMS magnet yoke during the solenoid discharge; Section 3 describes the measurements of the magnetic flux density inside the CMS coil and the measurements of the magnetic flux density in the CMS flux-return yoke blocks; there is a small discussion in Section 4; Section 5 contains a conclusion.

## 2. Materials and Methods

### 2.1. Developing a Technique for Measuring the Magnetic Field Inside the CMS Solenoid

The system of the NRM probes installed inside the superconducting solenoid uses the Metrolab Technology SA probes of model 1062, connected through a 2030 multiplexer to a programmable teslameter PT 2025 [13]. The schematic view of the 1062 NMR probe is shown in Figure 1a. There are three types of sensors used in the system: one probe (B) has an active volume made of a solid material containing a large amount of hydrogen to measure the magnetic induction in the range from 0.7 to 2.1 T; one probe (F) has a sealed glass tube containing heavy water ($D_2O$) to measure the magnetic field in the range from 1.5 to 3.4 T; four probes (A, C, D, E) have a similar NMR sample filled with $D_2O$ to measure the magnetic flux density in the range from 3 to 6.8 T. The magnetic field measurement interval is determined by the frequency of the high-frequency generator signal, which is from 30 to 90 MHz for the probe B, from 7.5 to 22.5 MHz for the probe F, and from 15 to 45 MHz for all other probes. This signal is fed to a radio frequency coil located on the active element, and with small perturbation of the strong measuring field by the modulation coil creates conditions for the NMR effect to appear in the active volume, when the oscillation frequency matched the frequency of the nuclear magnetic moment precession around the magnetic field lines and enhances the signal.

The modulation frequency of the measured magnetic field is from 30 to 70 Hz and is generated by an additional generator of a triangle signal applied to the modulation coil, the plane of which is located at an angle of 45° with respect to the direction of the measured magnetic flux density, which allows one to measure the transverse or axial magnetic field. Calibration of sensors in a known magnetic field makes it possible to bind the observed oscillation frequency with the value of the corresponding magnetic flux density of the measured field. The combination of the high-frequency and modulation signals provides an accurate magnetic field measurement with a resolution of 0.1 µT (1 Hz in frequency).

The active volumes of probes A and B are located at a radius of 2.9148 m from the solenoid axis and $Z = -0.006$ m from the CMS detector median $XY$-plane at azimuth angles of 44.9° and −135.1°, respectively. The active volumes of probes F and E are located at the same radius and $Z = +0.006$ m at azimuth angles −44.9° and +135.1°, correspondingly. Probe C measures the magnetic field at a point with coordinates ($X$, $Y$, $Z$) = (0.6425, 0.10517, −2.835) m, and probe D—at a point with coordinates ($X$, $Y$, $Z$) = (0.6425, 0.10517, 2.831) m; both sensors are located at the faces of the tracking system volume.

Teslameter PT 2025 is in the underground service cavern and is connected by three 64-m cables to the 2030 multiplexer located in the underground experimental cavern and connected with 5 NRM sensors by 30-m cables and with one by 35-m cable.

The range of magnetic flux density, $B$, measured with the NMR probes at the solenoid current values varying from 4 to 19.14 kA, covers the interval from 0.85 to 4.01 T. In this interval, the dependence of $B$ on the solenoid current is linear.

The precise measurement of the magnetic field in the cylinder volume of 1.724 m radius and 7 m long inside the CMS coil has been done in 2006 with a fieldmapper designed and manufactured at FNAL [8]. The fieldmapper comprised ten 3D B-sensors [14–16] developed at National Institute for Subatomic Physics (Nikhef, Amsterdam, the Netherlands) and calibrated at CERN to a precision of $3·10^{-4}$ at 4.5 T field [17–19]. Two NMR probes with the active volumes filled with $D_2O$ were used in addition to measure the field along the coil axis and at the largest radius of the measured volume.

The fieldmapper shown in Figure 1b inside the measured volume moved along the rails installed along the coil axis in the barrel hadron calorimeter, stopping at predefined points where two arms with B-sensors could be rotated through 360°, stopping at predefined angles where the magnetic field was sampled. The azimuth steps were 7.5° in magnitude. Steps along the coil axis were fixed to 0.05 m by a tensioned toothed Kevlar belt.

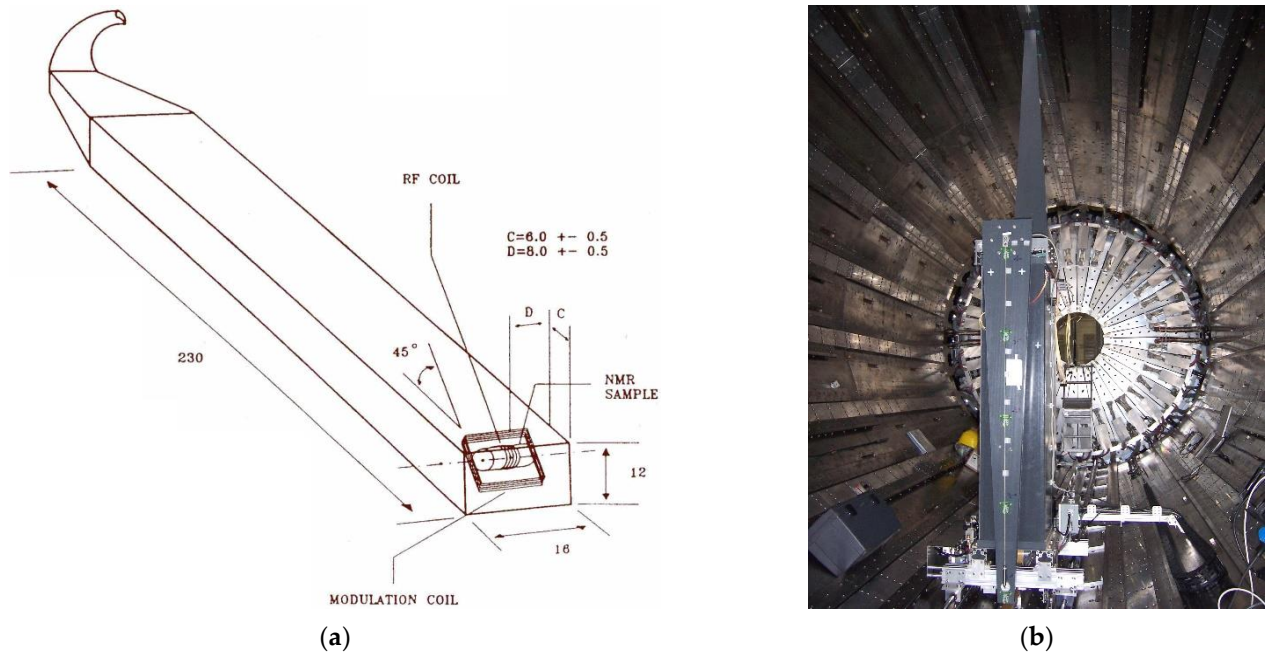

(**a**)          (**b**)

**Figure 1.** (**a**) Schematic view of a nuclear magnetic resonance probe. The probe external dimensions in mm (230 × 16 × 12), the position of an active volume (NMR sample) with a radio frequency (RF) coil, as well as the slope of a modulation coil, equal to 45° with respect to the probe axis, are shown. The NMR sample has a diameter of 4 mm and a length of 4.5 mm and is made of either a solid material containing a large amount of hydrogen or a sealed glass tube containing $D_2O$. The measured magnetic field direction can be transverse or axial; (**b**) Automated field-mapping machine [8] for measuring the CMS magnetic field, installed inside the barrel hadron calorimeter. A carriage made of aluminum alloy moving by steps of 0.05 m along the rails aligned with the *Z*-axis, a tower made of durable non-magnetic material, two propeller arms rotating by steps of 7.5° along the azimuth angle in the forward and backward directions, and five 3D B-sensors on the propeller arm viewed from the positive *Z*-coordinates are visible.

Each arm of the fieldmapper contained five 3D B-sensors located at radii 0.092, 0.5, 0.908, 1.316, and 1.724 m off the coil axis. The distance between the negative and positive arm B-sensors along the coil axis was 0.95 m.

Made of nonmagnetic materials, the fieldmapper used pneumatic power. The high-purity nitrogen gas flow was controlled with 24-V piezoelectric valves, the remote operation was performed via a programmable logic controller and operator's LabVIEW [20] console. The laser ranger was used for absolute *Z*-coordinate reference after unscheduled stops.

The alignment of the fieldmapper azimuth axle with respect to the CMS coil axis was performed with a precision better than 1.9 mrad. The read-out of the B-sensors was performed via the CANopen protocol [21,22].

*2.2. System for Monitoring the Magnetic Flux Density during the CMS Detector Operation*

The first stage of the CMS magnetic flux monitoring system consisted of 86 3D B-sensors: 17 high field B-sensors calibrated at 4.5 T magnetic field, and 69 low field B-sensors calibrated at 1.4 T field [17–19]. The B-sensor printed circuit board (PCB) is shown in Figure 2a. Three single-axis Siemens GaAs Hall sensors of type KSY44 [23] with dimensions of $3.2 \times 2.3 \times 0.6$ mm$^3$ each were glued on three orthogonal surfaces of a glass cube with dimensions of $4 \times 4 \times 2.4$ mm$^3$. The used Hall current was 230 μA, which led to small heat dissipation. Hall voltages were sampled by a 24-bits delta-sigma modulator to perform the signal analogue-to-digital conversion. To measure a temperature nearby the Hall probes a calibrated thermistor is connected to the cube, no thermostat is used. Three precision holes in support legs allow one to mount the PCB on the CMS detector parts. All the PCB analogue electronics resisted the measured field. An 8-byte ID-chip (DS2401 Dallas) on each B-sensor PCB helps to administrate large number of cards in the experiment area.

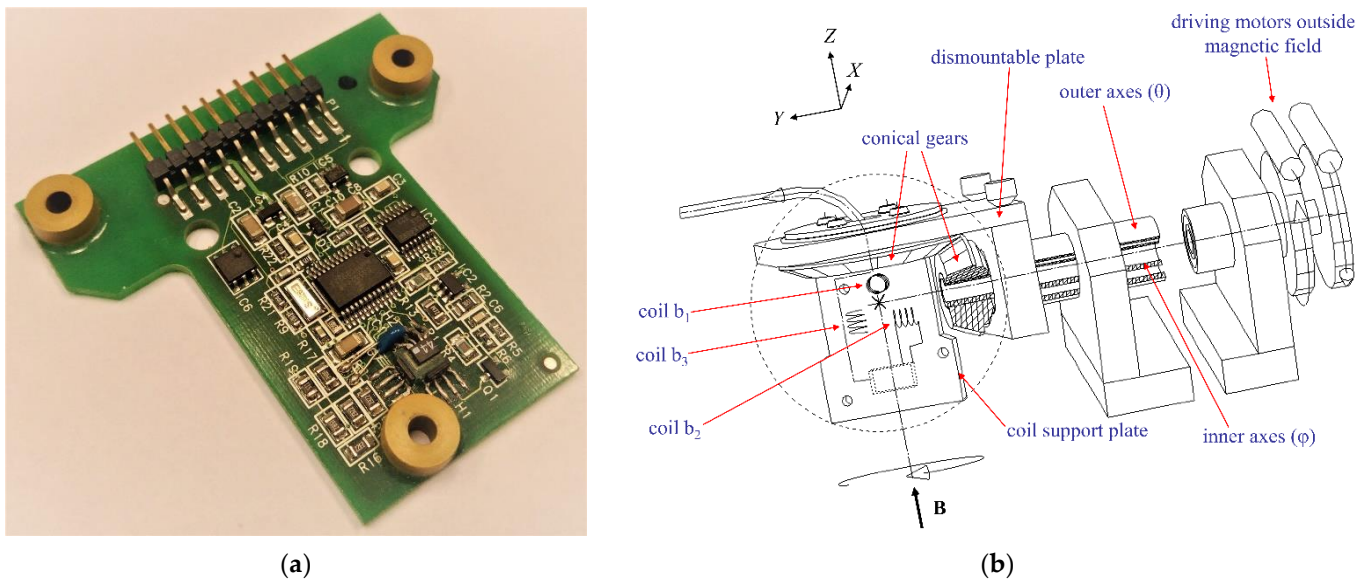

(**a**) (**b**)

**Figure 2.** (**a**) Hall probes on the B-sensor PCB. Each PCB contains three single-axis Siemens KSY44 Hall chips [23] which are glued to a glass cube of $4 \times 4 \times 2.4$ mm$^3$. The distance between the $b_1$ (at the cube top) and $b_3$ (at the H1 side) chip centers is 1.8 mm. The distance between the $b_1$ and $b_2$ (at the R18 side) chip centers is 2.6 mm. The B-sensors have an orientation error of about 1 mrad, and the relative orientation error of local $b_1$, $b_2$, $b_3$ measured fields is estimated to be approximately 0.2 mrad [24]. The analogue voltages from the Hall probes are simultaneously read out by a 24-bit ΔΣ-modulator; (**b**) The Hall probe calibrator scheme [17–19]. The local coordinate system *XYZ* is rotated with respect to the constant magnetic flux density vector ***B*** in two angular directions: a polar angle *θ* is counted between ***B*** and the *Z*-axis, and an azimuthal angle *φ* is counted between the projection ***B***·$sinθ$ and the *X*-axis. The rotations are performed with the calibrator outer and inner axis providing the rotations of the calibrator head in *θ* and *φ* directions, accordingly. To cover the full 4π space in the local reference frame, 6 turns of the outer axis and 5 turns of the inner axis in the opposite directions are needed. Four B-sensor PCB with the same orientation are mounted by two on each side of the coil support plate. Three coils measure the components $b_1$, $b_2$, and $b_3$ of ***B*** in the local coordinate system by the magnetic flux integration. The Hall probe voltages and the coil signals are sampled each 1/15 s and approximated then by the orthogonal spherical harmonics with a set of calibration coefficients at three values of ***B*** and two values of temperature.

A calibration of the 3D B-sensors to a precision of $5 \cdot 10^{-4}$ [10] has been performed by rotation of a package of four PCBs in a constant, homogeneous magnetic field at different absolute field values and temperatures [17–19]. A special calibration device shown in Figure 2b was designed and manufactured at CERN for this purpose. The calibrator rotated a small head in three constant values of the homogeneous magnetic flux density ***B*** in two orthogonal angular directions *θ* and *φ* with help of stepping motors located outside the magnetic field and a special transmission consisted of the long inner and outer axis and conical gears. The head was rotated in a thermostat box where a constant temperature was maintained by a Peltier thermoelectric element, and the cooling air flow was controlled by a fan. The middle plate of the head contained three orthogonal coils $b_1$, $b_2$, and $b_3$ to measure the magnetic flux density components by the magnetic flux integration in each coil in the head local reference frame shown in Figure 2b. On each side of the coil support plate, two B-sensor PCBs were installed.

The polar angle *θ* was counted in the head local coordinate system as an angle between the vector ***B*** and the *Z*-axis. The azimuthal angle *φ* was counted as an angle between the projection ***B***·*sinθ* and the *X*-axis. To complete the full 4*π* space with rotation of vector ***B*** in the local coordinate system, 6 turns of the head in *θ*-direction and 5 turns of the head in *φ* direction were needed. A signal/power supply cable winded by the outer (*θ*) axis but unwinded by the inner (*φ*) axes. At the end of the calibration, the cable had made only one turn. The trajectories of the magnetic flux density unit vector in the head local reference frame are shown in Figure 3 in the *XY*- and *YZ*-planes.

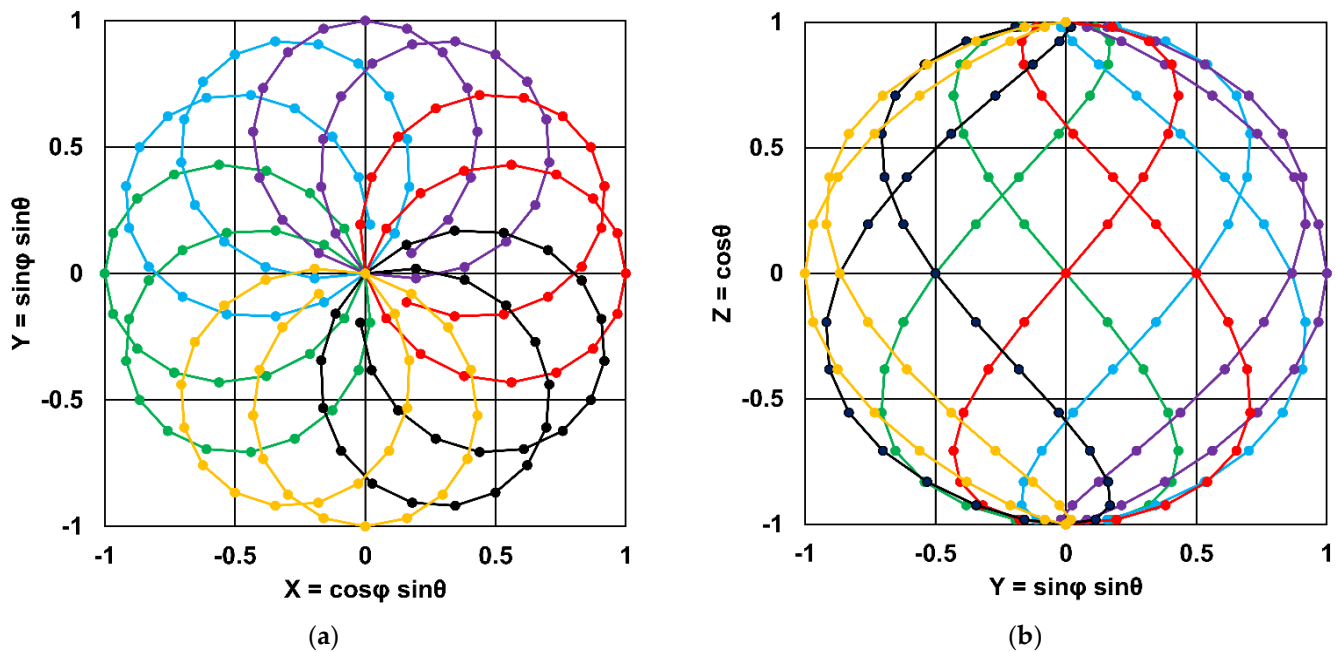

(**a**)  (**b**)

**Figure 3.** Trajectories of the magnetic flux density unit vector in the calibrator head local coordinate system: (**a**) in the *XY*-plane; (**b**) in the *YZ*-plane. Different colors correspond to six complete turns of the calibrator head with the outer axes. Markers denote the increments of 9.375° in azimuth *φ* and 11.25° in polar *θ* angles used to prepare the plot.

The Hall voltages $V_H$ induced in three single-axis Siemens KSY44 chips during the head rotation have been sampled each 1/15 s together with the coil signals and then were decomposed in orthogonal functions in the way as follows [18]:

$$V_H(|B|, t, \theta, \varphi) = \sum_k \sum_n \sum_l \sum_{m=0}^{l} c_{knlm} T_k(B) T_n(t) Y_{lm}(\theta, \varphi), \quad (1)$$

where $c_{knlm}$ are calibration constants, $Y_{lm}$ are spherical harmonics [25] of order *l*, *m* for spatial part, $T_n$ are Chebychev polynomials of the first kind [26] of order *n* for temperature dependence, $T_k$ are Chebychev polynomials of the first kind of order *k* for absolute field dependence, *θ* and *φ* are polar and azimuthal angles, *B* is absolute field value, and *t* stands for temperature.

A calibration procedure at given magnetic field value and given temperature required three minutes. For a given B-sensor the calibration has been performed for three values of the magnetic flux density (0.37, 0.885, 1.4 T for the low field B-sensors; 2.5, 3, 4.5 T for the high field B-sensors) and for two values of temperature (20 °C and 24 °C). The calibration constants have been stored in the database for each calibrated B-sensor and then were used to convert the measured Hall probe voltages into the magnetic flux density during the magnetic field measurements and monitoring.

*2.3. Developing a Flux Loop Technique of Measurements of the Magnetic Flux Density Inside the CMS Yoke Steel Blocks*

2.3.1. Concept of the Magnetic Flux Density Measurements in Steel with the Flux Loops

A procedure to measure the magnetic flux density inside the CMS yoke steel blocks had been proposed in 2000 [27] and assumed using the fast discharge of the CMS coil to induce voltages in flux loops installed around selected blocks of the CMS flux-return yoke. By sampling the voltage induced in any one loop and integrating the voltage waveform over the time of the discharge, the total initial flux in the loop can be measured. The voltage induced in any one flux loop is proportional to the number of turns in the loop. The average value of the magnetic flux density normal to the plane of the flux loop wound around the block is obtained by dividing the measured value of the magnetic flux by the known area enclosed by the loop and the number of turns in the loop.

The standard ramp up and ramp down time of the CMS magnet is approximately 4–5 h depending on the ramping rate, and the slow discharge time is about 19 h. With a rate of 1.5 A/s, the standard ramp down from an operating current of 18.164 kA to 1 kA takes 11442.7 s. Starting from a current of 1 kA the fast discharge is automatically triggered, and the current decay departs from a simple $L/R(t)$ decay of an inductor ($L$) into an external resistance ($R(t)$) changing with time ($t$), which requires another 3600 s. If the average initial magnetic flux density in the flux loop area of 1 m$^2$ is 1.5 T, then the initial magnetic flux $\Phi$ in the loop cross section is 1.5 Wb. Dropping this value to zero for 15042.7 s induces in one turn of the loop an electromotive force (EMF) voltage $V = \Delta\Phi/\Delta t$ of 0.0997 mV. With a loop made of 400 turns the induced voltage reaches 40 mV that requires a precise analogue to digital convertor (ADC) to separate this small signal from a noise. The number of turns was limited by the cross sections of the grooves in steel blocks used for the flux loop arrangements.

The fast discharge time constant is 190 s, which induces voltages with a much larger amplitude. Evidently, measuring the flux loop voltages during coil fast discharge provided the best opportunity to make the intended measurements. During normal operation of the CMS magnet fast discharge of the superconducting solenoid is only triggered by the detection of some abnormal operating condition, which, from a safety point of view, requires discharge of the coil as rapidly as could be achievable. To verify the proper performance of the magnet safety system, several manually triggered fast discharges have been performed in 2006 during commissioning of the CMS magnet [7] before it was lowered into the CMS underground experimental cavern. These fast discharges provided an opportunity to make the flux loop measurements of the magnetic field in the yoke steel using a simpler ADC.

To estimate the amplitudes of signals induced in the flux loops during the fast discharges, the magnetic flux variations in the proposed flux loop displacements have been calculated with the CMS magnet 3D model [12]. At nine discrete times (0, 50, 100, 125, 151, 176, 200, 251, and 306 s) during the simulated fast discharge shown in Figure 4a the magnetic flux has been calculated in the entire CMS detector volume. In the blocks of the barrel wheels and the endcap disks, the resulting magnetic flux density values were integrated over the areas enclosed by the flux loops. From the total flux enclosed by each flux loop, the average voltages induced in the loops by the flux changes between time intervals were calculated as shown in Figure 4b,c. The presence of eddy currents in the steel cores of the flux loops, and their effects on the induced voltages in the loops, have been ignored in the calculations presented in Figure 4b,c [12].

The conclusion of this modelling was that the voltages induced in the flux loops could be integrated with a good precision to obtain the magnetic flux and then the average magnetic induction in the selected cross sections of the CMS flux-return yoke.

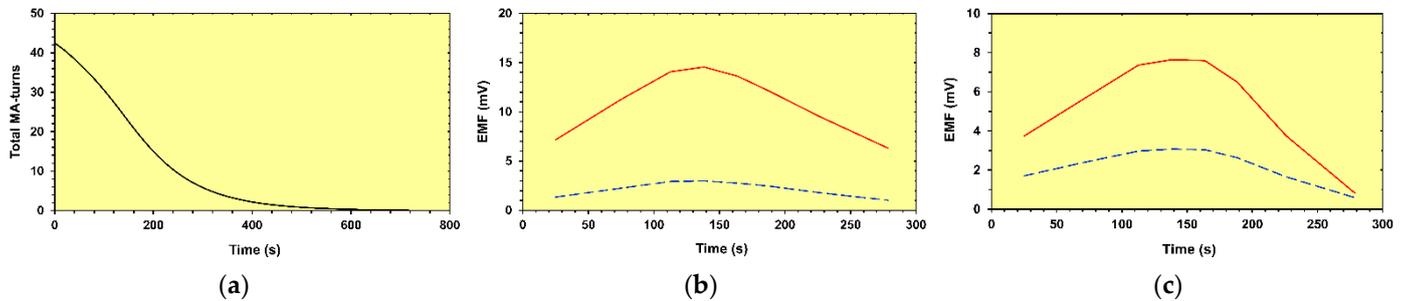

**Figure 4.** Modelled (**a**) CMS coil current fast discharge; (**b**) minimum (dashed) and maximum (solid) EMF voltages per one-turn flux loop on the blocks of the CMS barrel wheels; (**c**) minimum (dashed) and maximum (solid) voltages per one-turn flux loop on the 18° segments of the CMS endcap disks [12].

According to this study each flux loop was made of 7–10 turns of 45-conductor ribbon cable wound into a shallow groove of 30 mm wide and 12–13 mm deep machined into the peripheral surface of the steel block to be sampled. By connecting the two ends of the loop ribbon cable so that the individual conductors in the ribbon are offset by one conductor, the 315-450-turn flux loops were formed to encircle the selected parts of the yoke. As can be seen in the above-calculated EMF estimates shown in Figure 4b,c, voltages peaking to several volts could be induced in the multiple-turn flux loops.

2.3.2. Performance of a Special R&D Program to Model the Flux Loop Measurements

To verify if these voltages can be measured online and integrated offline over the entire CMS fast discharge with an accuracy of a few percent, a special R&D program was performed with several sample disks 127 mm in diameter and 12.7 or 38.1 mm thick made of the CMS yoke steel. Each sample disk was inserted between the poles of a test dipole magnet discharged from a maximum current of 320 A with a current shape similar to the shape of the CMS current generated by the solenoid fast discharge. The induced voltages were measured in a test flux loop mounted on the sample disk. To provide an equivalent variation in the magnetic flux, the number of turns in the test flux loop was larger than the number of turns in the CMS flux loops, and the duration of the test magnet discharge was shorter than the CMS fast discharge time.

The 994-turn test flux loop of 140.6 mm in average diameter was wound on a non-metallic bobbin and connected to the sampling circuitry in differential mode to reject common-mode noise [12]. The differential inputs of the ADC system were referenced to ground through 100 kΩ resistors.

The test magnet of GMW Associates Model 3474 was energized with Danphysik Model 8530 power supply equipped with General Purpose Interface Bus (GPIB) control interface, and the magnet charging and discharging were performed at several different rates under control by the same software used to sample the voltage on the test flux loop. The diameter of the test flux loop was chosen to fit within the flat portion of the pole tips of the test magnet. A test magnet model shown in Figure 5a was calculated with the TOSCA program [9] to interpret the data obtained from the test flux loop. The *B-H* curve for the test magnet pole tips and yoke were taken from the measurements of the CMS yoke steel samples.

Numerous sets of measurements were performed with the pole tip gap of the test magnet set to 12.7 mm and 44.45 mm [12].

In particular, the sample disks of 38.1 mm thickness made of different steel used in the CMS yoke were inserted in a 44.45 mm gap of the test magnet. In these studies, 3.175 mm air gaps between the samples and the pole tips of the test magnet were used to mount the Hall sensors on both sides of the disks at the centers of each side in the air gaps. The Hall sensors measured the axial magnetic flux density on the steel-air interface when the test magnet was fully energized, and the remanent field in the steel at the end of the discharge.

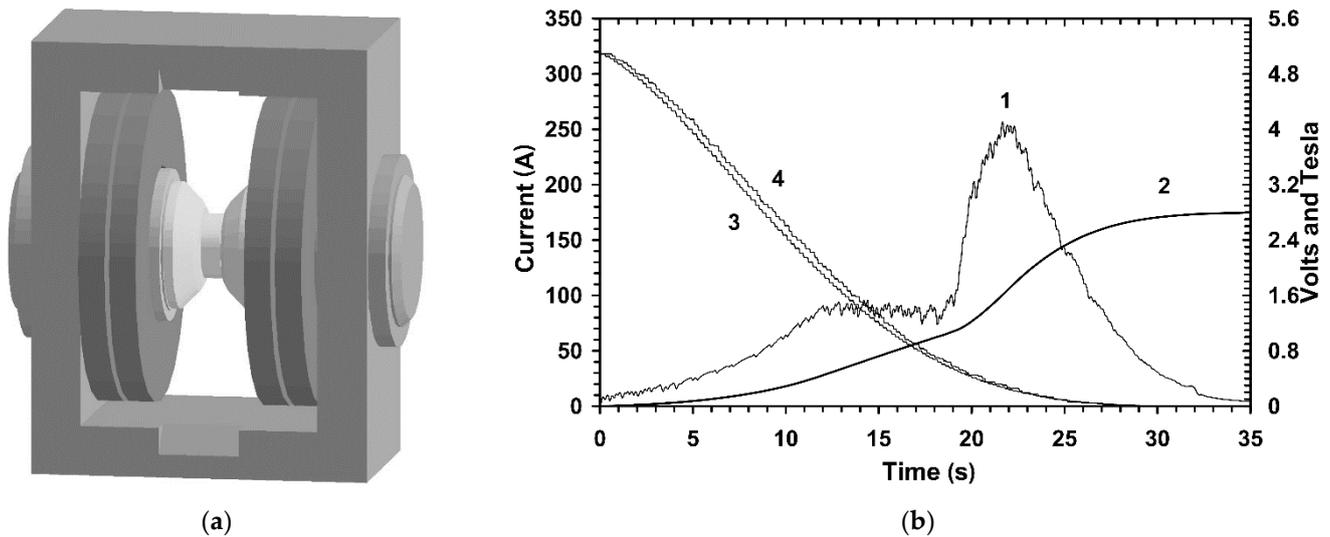

(**a**) (**b**)

**Figure 5.** (**a**) 3D model for the test magnet with the steel sample disk inserted between the pole tips [12]; (**b**) Induced voltage (curve 1) and the integrated magnetic flux density (curve 2) when the test magnet current ramped down from 320 A to zero during 32 s. Curve 3 shows the requested current from the control software. Curve 4 corresponds to the measured current read-back [12].

The TOSCA model predicted closely the flux density of 2.65 T in the 12.7 mm free air gap between the pole tips versus that measured by Hall probes (2.63 T) positioned on the surface of the pole tips when the test magnet was energized to full excitation. With the 12.7 mm thick steel sample disk inserted in the gap, the model predicted a field of 3.07 T in the center of the disk.

For the case with the 44.45 mm gap filled with the 38.1 mm steel disk spaced from the pole tips by two air gaps, the model predicted an axial field of 3.0 T in the center of the flat side of the disk when the test magnet is fully energized, whereas the Hall probes measured 2.9397 ± 0.0002 T. Based on this, a correction factor of 0.9799 was applied to other calculated values to be compared with the measured field values.

First, the test flux loop was inserted in the gap of 12.7 mm between the pole tips and the test magnet charged to full current of 320 A at a charge rate of 2.5 A/s. After a pause, the current was decreased at the same rate to zero. The voltage on the test flux-loop was sampled at 50 ms intervals (20 Hz sampling rate), and integrated offline by multiplying the average voltage in each time interval by the length of time interval. The measured flux changes from charging and discharging agreed within 2%.

Then, an aluminum disk 12.7 mm thick was placed in the flux loop and the assembly inserted between the pole tips. The behavior of the voltage induced in the test flux-loop was the same that excluded the substantial eddy currents in the metal sample disks.

The main studies were performed with two 38.1 mm thick sample disks made from the same steel as most of the CMS barrel yoke. Each was inserted into the test flux loop and spaced from the test magnet pole tips by air gaps of 3.175 mm. The charge-up of the test magnet was always at the rate of 2.5 A/s. Fast discharges have been studied with the overall discharge times of 32, 64, 128, 256, and 512 s with the shapes similar to the CMS fast discharge. Figure 5b shows the induced voltage and integrated magnetic flux density for the discharge time of 32 s. Before the charge-up and at the end of the discharge, the Hall probes measured the remanent fields in the air gaps $B_{r\,ch}$, and $B_{r\,dis}$, respectively. It was observed that these remanent fields increased for longer discharge times. It was 37 mT for 32 s discharges and increased to 59 mT for 512 s discharges. The eddy currents in the test magnet poles caused this effect, and it resulted in a long tail of the induced voltage after the current of the test magnet was set to zero. For all the discharges, this tail was measured during 70 s after $t = 0$ (after 32–512 s from the beginning of the discharge), where $t = 0$ represents the time when the current was requested by software control to become zero. In the case of the shortest discharge of 32 s, this tail contributed 1.8% to the integrated voltage. For the 512 s discharge, this contribution was 0.008%. The test magnet charge-up and discharge occurred as a series of small discrete steps in current visible in Figure 5b.

In eleven charge/discharge cycles of varying discharge times the sums $B_{r\,ch} + B_{i\,ch}$, and $B_{i\,dis} + B_{r\,dis}$ were investigated. In these sums $B_{i\,ch}$ is the magnetic flux density obtained from the magnetic flux integrated by the test flux loop during the charge-up of the test magnet, and $B_{r\,ch}$ is the remanent field measured by the Hall sensors before the charge-up began. The subscript "dis" denotes the same quantities measured during discharges of the test magnet (including 70 s after the current was ramped down to zero) with the Hall sensor value recorded after the discharge. Averaging the results from the eleven different cycles gave the values $\langle B_{r\,ch} + B_{i\,ch}\rangle = 2.8633 \pm 0.0018$ T for charging and $\langle B_{i\,dis} + B_{r\,dis}\rangle = 2.8583 \pm 0.0028$ T for discharging. The results agreed within 0.2%.

Taking the TOSCA calculations for the flux loop and scaling by the correction factor of 0.9799, a calculated magnetic flux density of 2.8726 T was obtained. This agreed with $\langle B_{r\,ch} + B_{i\,ch}\rangle$ within 0.3% and with $\langle B_{i\,dis} + B_{r\,dis}\rangle$ within 0.5%.

This special R&D program confirmed that the magnetic flux density in a steel object magnetized by an external source can be measured with good precision using a combination of the flux loops and Hall probes.

*2.4. Analysis of Eddy Current Distributions in the CMS Magnet Yoke during the Solenoid Discharge*

Right after the special R&D program described above was performed, sixteen 315–450-turn flux loops have been installed in azimuthal sector S10 at 270° of the central and two CMS negative side barrel wheels and another six 405–450-turn flux loops had been installed in azimuthal sector S10 of two CMS negative side endcap disks. To estimate the contribution of eddy currents to the voltages induced in the flux loops when the fast discharge of the CMS coil occurs, a special CMS magnet 3D model has been developed and calculated with Vector Fields' program ELEKTRA (Electromagnetic Analysis) [28].

Calculations with ELEKTRA, which utilizes a vector potential in the regions where the eddy currents are expected, are very CPU time consuming. To reduce CPU time to a reasonable amount, the CMS yoke was described in a simplified way, the number of finite element nodes in the models was reduced to a reasonable value, the time step varied from 6.25 to 25 s, and the number of output times in the transient analysis of the current decay following the drive function shown in Figure 4a did not exceed 15 [29]. To perform ELEKTRA analysis of eddy currents in the CMS yoke at 15 output times, 415 CPU hours on a 450 MHz processor machine was required. To meet the batch queue requirements and to vary the time step, the analysis restarted at 50, 100, 150, 200, and 300 s. To analyse the magnetic field distribution in absence of eddy currents at the same output times, 12.3 CPU hours were required.

The model shown in Figure 6 included the entire CMS superconducting coil at cryogenic temperature and a 1/24 segment of the yoke that was then rotated and reflected in the OPERA-3d (an OPerating environment for Electromagnetic Research and Analysis) [30] postprocessor analysis to obtain the full description of the CMS yoke. This 30° azimuthal segment of the yoke was described as two and one-half three-layered barrel wheels, a small nose disk, and two thick endcap disks. Neither the connection brackets between the barrel layers nor the azimuth gaps in the CMS barrel wheels were modelled. The thin endcap disks and ferromagnetic parts of the CMS forward hadronic calorimeter were also omitted.

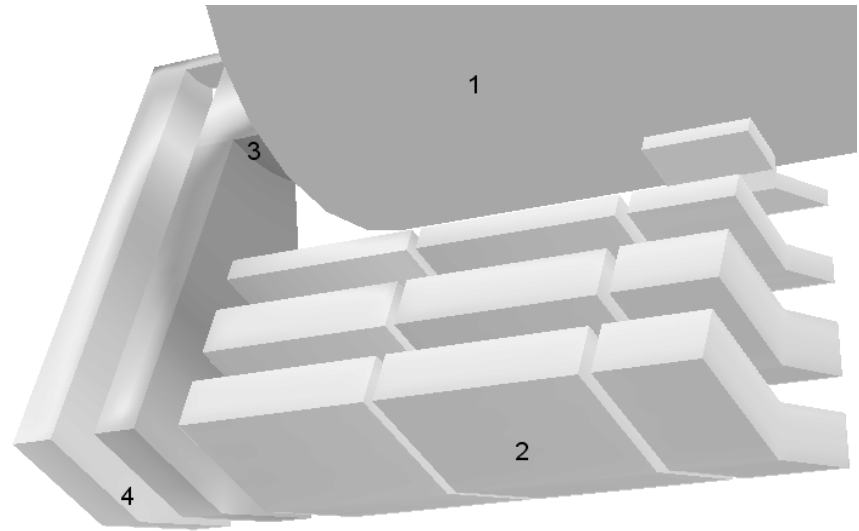

**Figure 6.** ELEKTRA model used for the yoke eddy current calculation [29]. CMS coil (1), the yoke sectors of the barrel wheels (2), nose disk (3), and two endcap disks (4) are presented in the model.

Different magnetic and electrical properties of materials were used to describe three different regions of the yoke: the tail catcher (TC, an additional inner layer of the central barrel wheel) and the first full-length thin barrel layer (L1) (region 1); second (L2) and third (L3) thick barrel layers (region 2); the nose and endcap discs (region 3).

A vector potential was used in all three regions. The electrical resistivity of construction steel used in calculations in regions 1, 2, and 3 was equal to 0.18, 0.15, and 0.165 µΩ, respectively.

The calculations of eddy currents in the CMS yoke were performed with ELEKTRA at 0, 25, 50, 100, 125, 150, 175, 200, 250, 300, 350, 400, 500, 600, and 700 s from the start of the simulated CMS fast discharge. At the same output times another ELEKTRA analysis was done with the model, which assumes an infinite electrical resistivity and total scalar magnetic potential instead of vector potential in all regions of the yoke.

The maximum eddy current density was investigated in the 22 flux loop steel cores. The calculation indicated that the maximum eddy currents in the yoke barrel cross sections arrived at 140 s after the beginning of the discharge, where the derivative of the current with respect to time reached an extreme. The eddy currents in the yoke endcap disk cross sections reached the maximum approximately 20 s later [29].

In the TC flux loop core the maximum eddy current density was 2.59 kA/m². In the L1 flux loop cores the maximum eddy current density varied from 4.16 to 12.9 kA/m². In similar cross sections of the L2 barrel layer, the maximum eddy current density varied from 5.14 to 12.5 kA/m². In the cross sections of the L3 barrel layer, the maximum eddy current density varied from 5.42 to 7.38 kA/m².

In the flux loop cores of the first endcap disk D−1, the maximum eddy current density varied from 27.12 to 51.98 kA/m² and, in the cross sections of the second endcap disk D−2, the maximum eddy current density varied from 11.21 to 17.52 kA/m².

To investigate if these values of the eddy current density change the magnetic flux, and thus, the average magnetic flux density in the yoke cross sections enclosed by the flux loops, the average voltages induced in the flux loops by the magnetic flux changes between time intervals were calculated as shown in Figure 7a,b.

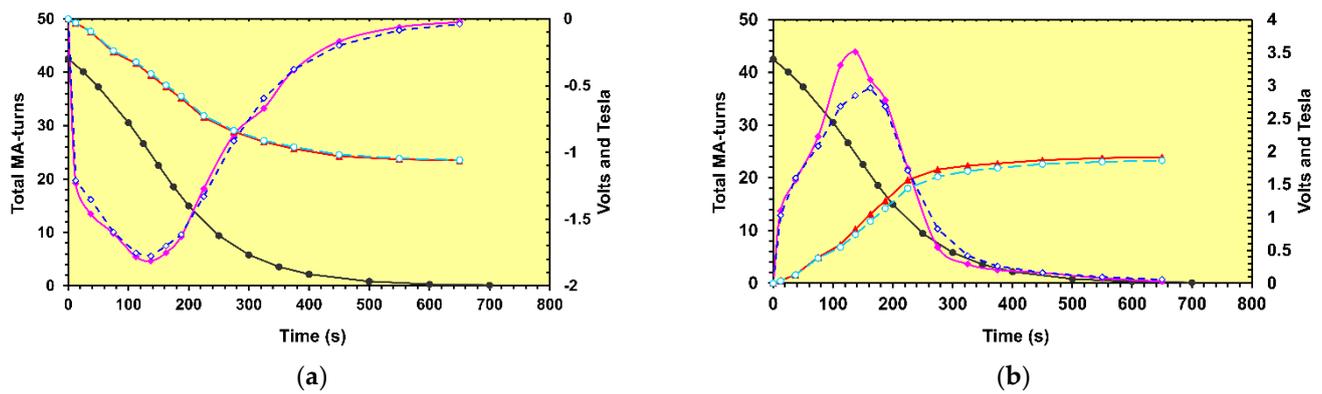

(a) (b)

**Figure 7.** (**a**) Voltages calculated in the first flux loop on the L2 layer of the external barrel wheel when the eddy currents with realistic electrical resistances (dotted blue line with open diamonds) and infinite resistances (smoothed solid magenta line with filled diamonds) are modelled during the current fast discharge (black solid line with black circles). The dashed light blue line with open circles represents the result of voltage integration when the eddy currents exist. The solid red line with filled triangles displays the result of voltage integration in the model with eddy currents suppressed. The difference between two integrated magnetic flux densities is within 0.3%; (**b**) Voltages calculated in the middle flux loop on 18° segment of the D−2 endcap disk when eddy currents from realistic electrical resistances (dotted blue line with open diamonds) and eddy currents suppressed by infinite resistances (smoothed solid magenta line with filled diamonds) are modelled during the current fast discharge (black solid line with black circles). The dashed light blue line with open circles represents the result of voltage integration when eddy currents exist. The solid red line with filled triangles displays the result of voltage integration when eddy currents are suppressed. The difference between two integrated magnetic flux densities is within 2.8%.

The voltages obtained in both models were integrated by multiplying the average voltage in each time interval by the length of time interval. The time integrals of the voltages are the total flux changes in the flux loops. The obtained flux values were renormalized to magnetic flux density using the areas of the flux loops and the numbers of turns in the flux loops.

The expected average eddy current contributions were found as follows: 0.22% ± 0.89% in the flux loop cores on the barrel wheels; −0.83% ± 2.42% in the flux loop cores on the endcap disks; and −0.067% ± 1.55% in all the yoke cross sections enclosed by the flux loops. A minus sign indicates that the value of the average magnetic flux density integrated in the model with eddy currents is less than the same value in the model without eddy currents.

These contributions lay well within the expected uncertainties of 2–3% anticipated in the flux coil measurements of the average magnetic flux density in steel elements of the CMS yoke, as was determined in the special R&D program [12].

## 3. Results

### 3.1. Measurements of the Magnetic Flux Density Inside the CMS Coil

Measuring the inner coil volume with the two-arm field-mapping machine was done at five different values of the central magnetic flux density $B_0$: 2.02, 3.02, 3.52, 3.81 (twice), and 4.01 T. At each of these values, the magnetic flux density was mapped at 5 radial distances in 141 azimuth planes with 48 azimuth angles; thus, the full number of the space points mapped was 33840. The radial distances of the 3D Hall probe locations at each arm of the fieldmapper were 0.092, 0.5, 0.908, 1.316, and 1.724 m off the coil axis coinciding with the CMS Z-axis. The distance between the negative and positive arm Hall probes along the coil axis was 0.95 m. Making 19 steps along the coil axis, the fieldmapper delivered the Hall probes of one arm in the same Z-position where the Hall probes of another arm were before and vice versa. Thus, the central part of the volume in Z-range of ±2.55 m with respect to the coil middle plane was mapped twice in the same pass of the fieldmapper through the volume with B-sensors of positive and negative arms. The difference between the magnetic flux density $B$ measured in the same point with the B-sensor of positive and negative arm did not exceed 1 mT.

In Figure 8a, the magnetic flux density $B$ measured at 4.01 T central field near the coil axis with Hall probes located at radii 0.092 m is displayed without any corrections for the B-sensor misalignment. This plot demonstrates the high quality of the measurements and shows no variations of $B$ with the azimuth angle. The general precision of measurements is 0.07%.

In Figure 8b, the magnetic flux density measured for the central field of 4.01 T at the radius of 0.092 m with the B-sensors is compared with values calculated with the CMS model version 1103_090322 [1]. The model was described in a half cylinder with a radius of 30 m and a length of 70 m in the configuration corresponded to the CMS magnet commission of 2006 [7]. The measurements differ from the calculated values by 2.1 ± 2.0 mT, on average, for the B-sensor located on the negative fieldmapper arm and by 1.4 ± 1.6 mT, on average, for the B-sensor located on the positive arm. For both B-sensors the measurements and calculations were averaged for each Z-coordinate over the full range of azimuth angle. The typical standard deviation of the measurement averaging is $4 \times 10^{-5}$.

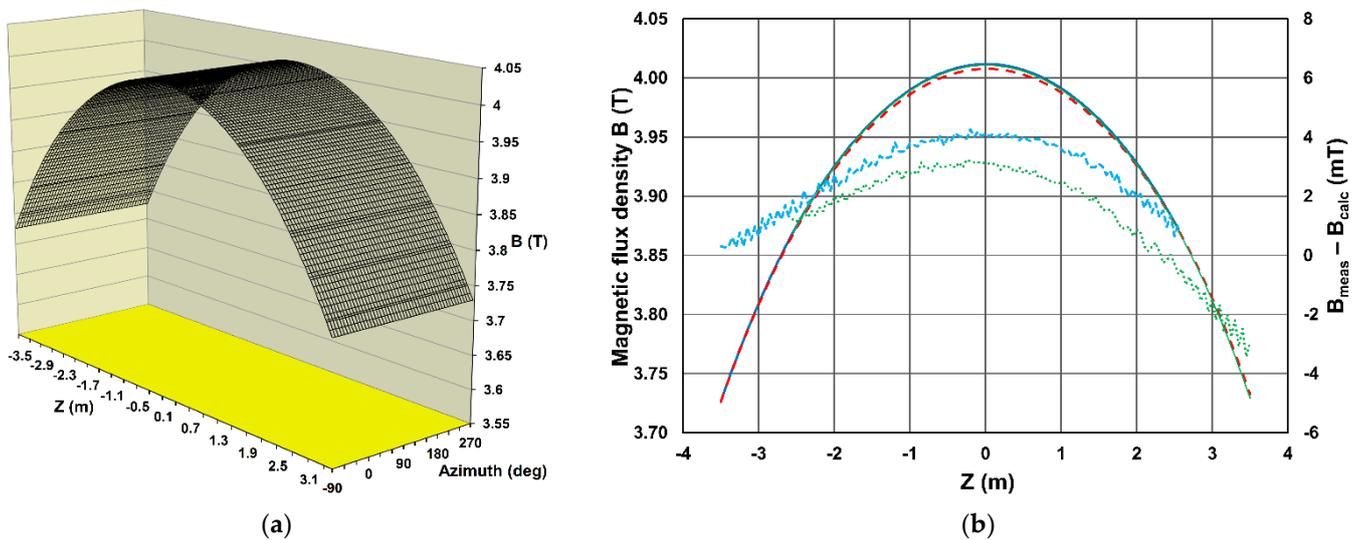

(a) (b)

**Figure 8.** (**a**) Magnetic flux density measured [8] at a radius of 0.092 m along the coil axis in the range of ±3.5 m with respect to the coil middle plane for full azimuth coverage; (**b**) Comparison [10] of the measured (smooth curves) and modelled (dashed curve) values of the magnetic flux density (left scale) averaged over the full azimuth angle range. The measurements have been performed with two B-sensors located at a radius of 0.092 m with respect to the coil axis on the negative (thick smooth curve) and positive (thin smooth curve) fieldmapper arms, respectively. The differences between the measured and calculated values (right scale) are shown by square and round dots, respectively.

In addition, at central field of 4.01 T the magnetic flux density was measured with two NMR probes moved with the fieldmapper along the coil axis and along the maximum radius of 1.724 m in horizontal plane. The field gradient and noise conditions allowed performing the measurements on the axis in Z-range from −1.675 to 3.025 m and the measurements at the maximum radius in Z-range from −1.767 to 2.583 m. At the maximum radius, the measurements with negative and positive arm B-sensors were also performed in the full Z-range of ±3.5 m.

In Figure 9a, the measurements done with the NMR-probe on the coil axis and the values calculated with the model version 16_130503 [1] are compared. The measurements differed, on average, from the calculated values by 2.4 ± 1.2 mT.

In Figure 9b, the comparisons between the measurements done with the NMR probes and B-sensors at the maximum radius of 1.724 m and the values calculated with the model version 16_130503 are presented. The measurements differ, in average, from the calculated values by 2.9 ± 1.0 mT for the NMR probe, by 2.0 ± 2.0 mT for B-sensor located on the negative fieldmapper arm, and by 1.7 ± 2.2 mT for B-sensor located on the positive arm of the fieldmapper. Both NMR probe and B-sensor measurements consist well.

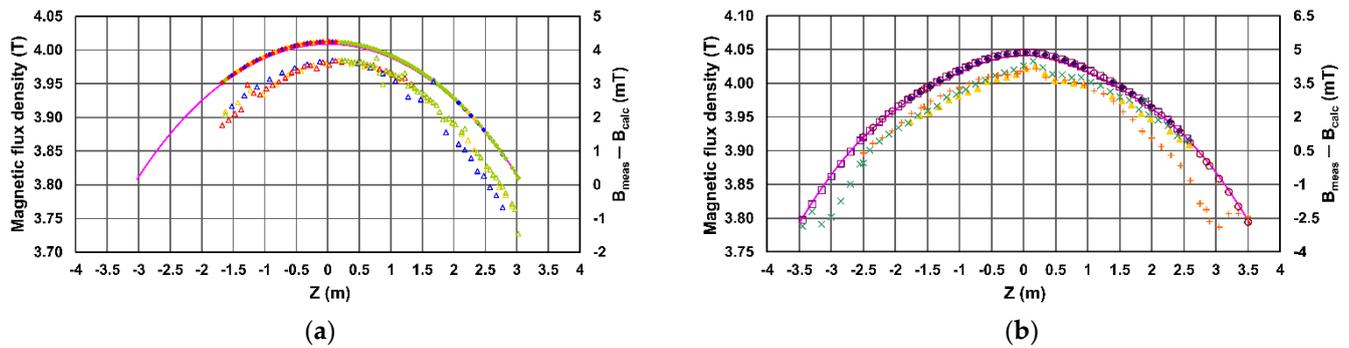

**Figure 9.** (**a**) Magnetic flux density (left scale) measured [8] with the NMR probe (rhombs) along the coil axis in the range of Z-coordinate from −1.675 to 3.025 m and calculated (smooth line) with the magnet model version 16_130503 [1]. The difference between the measurements and calculations (right scale) is shown by triangles. Different colors correspond to 4 sets of measurements; (**b**) Magnetic flux density (left scale) measured [8] with the NMR probe (rhombs), B-sensor of negative arm of the fieldmapper (open squares), and B-sensor of positive arm (open circles) all located at the radius of 1.724 m. The measurements are compared with the modelled values (smooth lines). The differences between the measured and calculated values (right scale) are shown by filled triangles, slanted and right crosses, respectively.

These comparisons show an excellent agreement between the modelled and measured magnetic flux density at central field of 4.01 T. In addition, the 3D model with one missing turn explains well the very small magnetic field asymmetry observed in the measurements. No axial shift of the coil with respect to the yoke was required to fit the measurements and the calculations. This conclusion was confirmed by the coil alignment and the coil cryostat position measurements.

To cross-check the latest CMS magnet model version 18_170812 [1], a comparison of the modelled magnetic flux density with the measurements done with four NMR probes and four 3D B-sensors installed inside the solenoidal volume was made at the operational current of 18.164 kA [30].

Two NMR probes are located near the coil middle plane at the Z-coordinates of ±0.006 m and radii of 2.9148 m; another two probes are installed on the CMS tracker faces at the Z-coordinates of −2.835 and +2.831 m and radii of 0.651 m. Four 3D B-sensors are located on the CMS tracker faces at the Z-coordinates of −2.899 and +2.895 m and radii of 0.959 m. The averaged precision of the NMR-probe measurements is $(5.2 ± 1.3) \times 10^{-5}$ T and that of the 3D B-sensors is $(3.5 ± 0.5) \times 10^{-5}$ T. The averaged relative differences between the magnetic flux density modelled and measured values were $(-5.6 ± 1.7) \times 10^{-4}$ at the NMR-probe locations, and $(-2.4 ± 4.0) \times 10^{-4}$ at the 3D B-sensor locations. These close results verify that the latest CMS magnet 3D model provides a good description of the magnetic flux distribution inside the superconducting solenoid volume.

*3.2. Measuring the Magnetic Flux Density in the CMS Flux-Return Yoke Blocks*

3.2.1. Flux Loop and B-Sensor Measurement System Description

Basing on the idea expressed in 2000 [27] and developed later [12,29], a system of the flux loops and 3D B-sensors was designed to measure the magnetic flux density in selected blocks of the CMS magnet flux-return yoke in the azimuthal sector S10 at 270°. An arrangement of this system is shown in Figure 10.

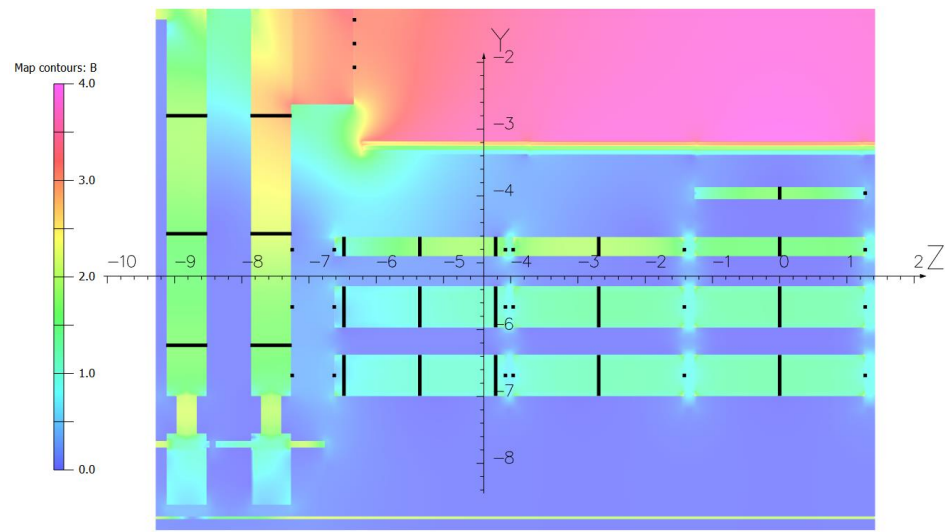

**Figure 10.** The magnetic flux density distribution in the longitudinal sections of the CMS detector. The colour scale is from zero to 4 T with a unit of 0.5 T. The black lines display twenty-two flux loop cross sections. The black squares denote the projections of the 3D Hall probe positions to the vertical *YZ*-plane. The values along the vertical *Y*-axis and horizontal *Z*-axis are presented in meters and are counted from the centre of the superconducting solenoid.

The system comprises 16 flux loops wound around 10 steel plates of the barrel wheels W0 (central), W−1 (adjacent to central), and W−2 (external wheel), 6 flux loops wound around the 18° segments of the endcap disks D−1 (the first) and D−2 (the second from the barrel yoke), 12 B-sensors installed on the inner surfaces of both nose disks, 31 B-sensors located in the air gaps on the faces of the barrel wheels W0, W−1, and W−2, and 18 B-sensors installed on the surfaces of the endcap disks D−1 and D−2. Another 18 B-sensors are installed symmetrically on the surfaces of the endcap disks D+1 and D+2. Finally, additional 3 B-sensors are installed in the azimuthal sector S4 located at 90° on the positive side of the D−2 endcap disk at the azimuth angle of 92°. The barrel wheel B-sensors were covered with special protection boxes made of G10 epoxy fiberglass plates; the endcap disk sensors were installed inside the special rectangular aluminum tubes with help of spring-loaded flexible strips; and the nose B-sensors were installed in two paralleled channels on each disk inside the boron polyethylene shield with help of special strips from the same material.

The Hall probes of the B-sensors were located in the air near the surfaces of the yoke steel blocks. At the interface between the two regions with different permeability, the normal components $B_n$ of the magnetic flux density satisfies the continuity conditions [31,32]; thus, at the air-steel interface, the normal flux density component measured by the B-sensor in the air is equal to the normal component $B_n$ inside the steel block. The flux loops located at the barrel wheels as shown in Figure 10 measure the $B_z$-component. In the analysis of measurements, the magnetic flux density was calculated with the CMS magnet 3D model in the areas where the measuring devices were located on the flux-return yoke. In addition to the flux loops, the magnetic flux density was also measured with the Hall probes installed 5 mm off the barrel wheels faces and 18 mm off the first endcap disk surface. The axial Z-coordinates of the Hall probes, also shown in Figure 10, were 1.273, −1.418, −3.964, −4.079, −6.625, and −7.251 m. These probes were aligned in rows at the vertical *Y*-coordinates of −3.958, −4.805, −5.66, and −6.685 m [33] on two sides of the magnet yoke: the near side toward the LHC center (positive *X*-coordinates) and the far side opposite to the LHC center (negative *X*-coordinates). With respect to the vertical plane at the *Y*-coordinate of −4.805 m, the rows were shifted by −0.56 m at the far side, and by 0.795 m at the near side of the yoke. The *X*-coordinates of the Hall probes were within the flux loop areas but near the edges of the flux loops. With *Y*-coordinate the *X*- coordinates of the probes followed the lines tilted by ±15 degrees with respect to the CMS vertical plane. Twelve 3D B-sensors installed 5 mm off the positive nose disk surface and 8 mm off the negative nose disk surface, as shown in Figure 10, were also used in the analysis and measured the magnetic flux density at radii from 1.373 to 2.082 m off the coil axis

The flux loops on the barrel wheel blocks were wound in 16 grooves of 30 mm wide and 12 mm deep on the TC block and on three barrel wheel layers L1, L2, and L3 at the Z-coordinates of 0, −2.691, −4.244, −5.352, and −6.48 m. The flux loops on the endcap disks D−1 and D−2 were wound in 6 grooves of 30 mm wide and 13 mm deep at the Y-coordinates of −2.8, −4.565, and −6.235 m. The loops have 7–10 turns of the 0.635 pitch flat ribbon cable reduced from fifty AWG 30 wires to 46 wires, 45 of which are used to form each loop. A special PCB shown in Figure 11 has been designed to form each flux loop by connecting 45 wires on both ends of the cable using 3M™ 1.27 mm pitch wiremount sockets, boardmount right angle plugs, and a special scheme to offset the individual conductors in the ribbon by one conductor at the ends of the cable. Two free wires at the ends of the cable related to the differential analog input of the AD-USB readout module from Measurement Computing [34]. The modules were attached by the USB cables to two network enabled AnywhereUSB®/5 hubs [35] connected to a PC through 3Com® OfficeConnect® Dual Speed Switch 5 [36] sitting on the network cable.

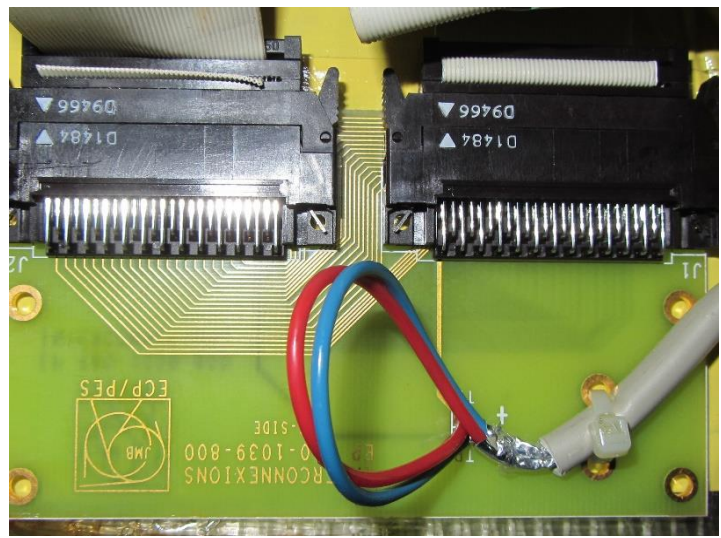

**Figure 11.** Double layer PCB to form the flux loop by connecting forty-five AWG 30 wires on both ends of the 0.635 pitch flat ribbon cable using 3M™ 1.27 mm pitch wiremount sockets, boardmount right angle plugs, and a special scheme to offset the individual conductors in the ribbon by one conductor at the ends of the cable. The flux loop relates to the readout AD-USB module by twisted pair screened cable connected to the PCB in two termination points: TP1 (red wire) and TP2 (blue wire). The PCB jack J1 is on the right side and jack J2 is on the left side.

To read out the voltages induced in the flux loops during 2006 magnet commissioning [8,11,37], the 12-bit USB-1208LS AD-USB modules from Measurement Computing were used. In 2013/2014 the readout system of the flux loop voltages was upgraded to replace these modules with new 16-bit USB-1608G modules from the same manufacturer. This replacement allowed measurements of readout voltages with a precision of 0.15 mV compared with a precision of 2.44 mV with the 12-bit modules. The new 16-bit readout gives a resolution of 0.75% at a typical amplitude of 20 mV. The local ethernet network cable of 90 m was replaced in 2013/2014 by the shielded optical faber cable of 100 m supplied with two Magnum CS14H-12VDC Convertor Switches on both ends. This modification allowed one to perform the flux loops measurements during the CMS magnet standard ramps down with the current discharge speed as low as 1–1.5 A/s with acceptable accuracy [30,38].

In the measurements, the average magnetic flux density components $B_i$ ($i = z, y$) orthogonal to the flux loop cross sections were obtained as $B_i = \Phi/A$, where the magnetic flux $\Phi$ was calculated by the integration of the signal voltages over the total time of the measurement [12]; $A$ is an area covered by the flux loop. The flux loop areas have been described considering position of each individual turn of the flat cable conductor and vary from 122 to 642 m². This approach reduces a systematic error arising from the flux loop conductor arrangement to ±3.6% on average in comparison with earlier estimations of the flux loop areas [8,11,37]. The flux loop area $A$ is calculated by the following expression: $A = N \times (a + c) \times (b + c) + d$, where $N$ is the total number of the flux loop wire turns, $a$ and $b$ are the width and height of the fifth turn of the flat ribbon cable, and $c$ and $d$ are small constant terms dependent on the number of turns $N$.

The calculations performed with the CMS magnet model have shown that the magnetic flux density is quite uniform in the flux loop cross sections.

3.2.2. Measurements of the Magnetic Flux Density in the Steel Yoke Blocks

Several attempts have been made to compare the results of the 2006 flux loop measurements with the 3D model calculations [8,11,33,37]. In these comparisons the main problem, that was not resolved completely, is a difficulty to estimate the eddy current contributions to the induced voltages in a correct way. The aluminium conductor around the superconducting cable was not included into the eddy current simulations [29]; thus, this unknown contribution of the eddy currents can overestimate the integrated magnetic fluxes and the averaged magnetic flux densities in the flux loop cross sections. In Figure 12, two examples of the real shapes of the voltages induced in the flux loops are displayed. From a comparison of Figure 12a with Figure 7a, the eddy current contribution can be estimated at the level of 4.6% as into the integrated magnetic flux, as into the integrated magnetic flux density. This is larger than it has been predicted in the eddy current analysis done with ELEKTRA program [29]. Both plots are prepared for the same flux loop cross sections, same numbers of turns and the same integration time.

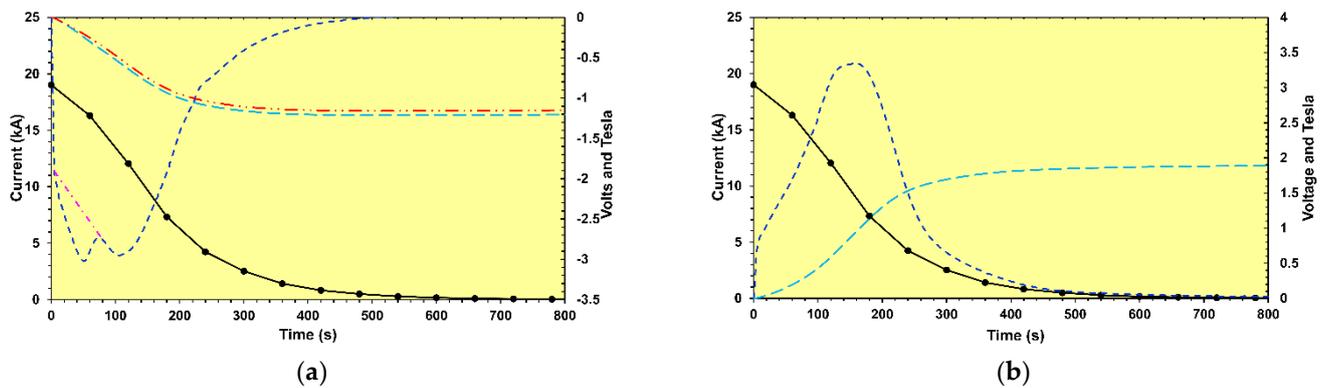

**Figure 12.** Voltages (smooth lines) induced in the flux loop of the W−2 barrel wheel second layer L2 (**a**), and in the middle flux loop of the D−2 endcap disk (**b**) in the 2006 magnetic field measurements. The integrated flux densities (dashed lines) and the fast discharge of the coil current from 19.14 kA (dotted lines) are also shown. In (**a**) the dashed-dotted line cuts the contribution of the eddy currents into the barrel wheel flux loop voltage. The dashed-double dotted line shows the integrated magnetic flux density without the eddy current contribution.

The best result of comparison of the 2006 flux loop measurements with the CMS magnet 3D model calculations is as follows [33]: the differences between the calculated and measured values of the magnetic flux density are of 0.59% ± 7.41% in the barrel wheels and −4.05% ± 1.97% in the endcap disks at the maximum current of 17.55 kA. These differences are of 1.41% ± 7.15% in the barrel wheels and −2.87% ± 2.00% in the endcap disks at the maximum current of 19.14 kA. The error bars of the magnetic flux density measured with the flux loops are of ±8.55% and include the errors in the knowledge of the flux loops geometries and the errors of the measured magnetic fluxes. The error bars of the 3D B-sensor measurements included into analysis are ± (0.025 ± 0.015) mT at the current of 17.55 kA and ± (0.012 ± 0.001) mT at the current of 19.14 kA.

An upgrade of the flux loop measurement system carried out in 2013/2014 allowed one to perform in 2015–2018 a new set of measurements [30,38]. To induce the voltages in the flux loops, seven linear discharges from a current of 18.164 kA to 0 kA were performed as shown in Figure 13a.

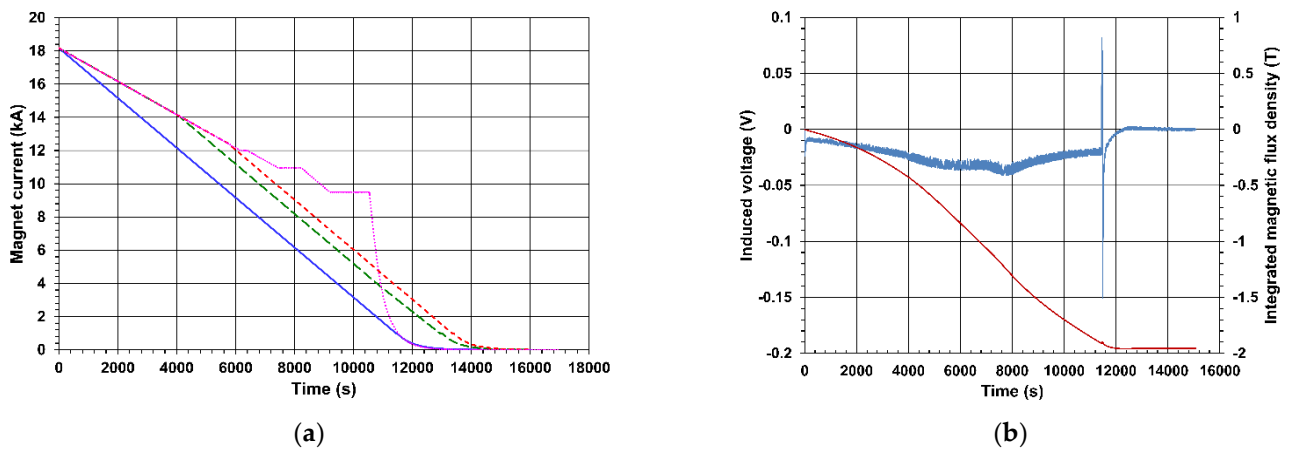

(a)  (b)

**Figure 13.** (**a**) CMS magnet current discharges from 18.164 to 0 kA made on July 17 and 18, 2015 (blue smooth line), September 21 and 22, 2015 (green dashed line), September 10, 2016 (red short-dashed line), and November 30, 2017 (magenta dotted line) [30]; (**b**) Induced voltage (left scale, noisy curve) and the integrated average magnetic flux density (right scale, smooth curve) in the cross section at $Z = -2.691$ m of the first layer block of the W−1 barrel wheel [30].

The first discharge, on July 17 and 18, 2015, was made with a constant current ramp down rate of 1.5 A/s to a current of 1 kA and, after a pause of 42 s, the fast discharge of the magnet was triggered manually to end the ramp down. The measurements of the voltages induced in the flux loops (with maximum amplitudes of 20–250 mV) were integrated over 15061.5 s in the flux loops located on the barrel wheels and over 15561.5 s in the flux loops located on the endcap disks [38]. The typical induced voltage in the first magnet ramp down, together with the integrated average magnetic flux density, is shown in Figure 13b. The rapid maximum and minimum voltages at 11,445 s correspond to the pause in the ramp down at a current of 1 kA and subsequent transition from the standard ramp down to the fast discharge of the magnet on an external resistor.

The second magnet discharge, on September 21 and 22, 2015, was performed with two constant ramps down rates: 1 A/s to a current of 14.34 kA (a central magnetic flux density $B_0$ of 3 T), and 1.5 A/s to a current of 1 kA.

The third magnet discharge, on September 10, 2016, was similar; however, the current at which the rate transitioned from 1 to 1.5 A/s was 12.48 kA. Changing the current rates was required by the cryogenic system operational conditions. In both these magnet ramp downs, the fast discharges were triggered from a current of 1 kA, and the offline integration of the induced voltages was performed over 17 000 s [30].

Another four discharges were performed using the second discharge scheme.

In Figures 14 and 15, the measured values (filled markers) of the magnetic flux density versus $Z$- and $Y$-coordinates are displayed and compared with the field values computed by the CMS magnet model (open markers) at the operational current of 18.164 kA. The lines shown in Figures 14 and 15 represent the magnetic flux densities modelled along the lines across the $XY$-coordinates of the B-sensors those are from 0.155 to 1.325 m away of the flux loop central $XY$-coordinates.

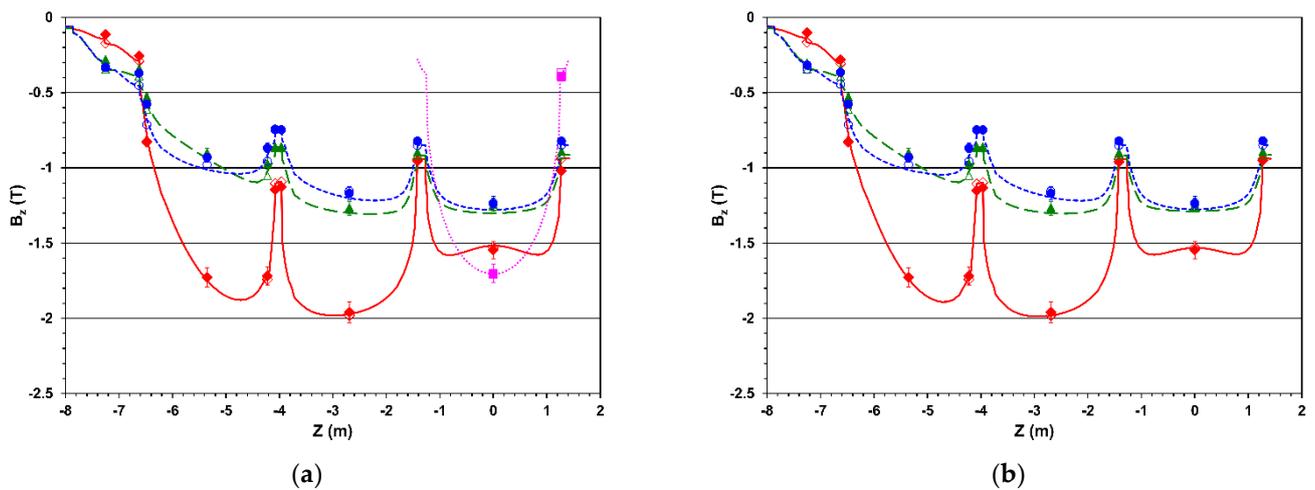

(**a**) (**b**)

**Figure 14.** Axial magnetic flux density measured at $B_0$ of 3.81 T (filled markers) and modelled (open markers) versus the Z-coordinate (**a**) in the TC (squares), and the L1 (diamonds), L2 (triangles), and L3 (circles) barrel layers at the yoke near side and the Y-coordinates of −3.958 m (dotted line), −4.805 m (solid line), −5.66 m (dashed line), and −6.685 m (short-dashed line); (**b**) in the L1 (diamonds), L2 (triangles), and L3 (circles) barrel layers at the yoke far side of and the Y-coordinates of −4.805 m (solid line), −5.66 m (dashed line), and −6.685 m (small dashed line).

The comparisons averaged over a set of seven measurements gave the following differences between the modelled and measured values of the magnetic flux densities in the flux loop cross sections: 4.1% ± 7.0% in the barrel wheels and −0.6% ± 2.7% in the endcap disks.

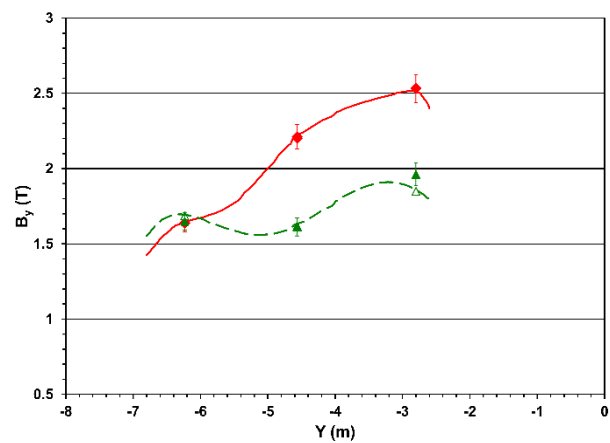

**Figure 15.** Radial magnetic flux density measured at $B_0$ of 3.81 T (filled markers) and modelled (open markers) versus the Y-coordinate in the D−1 (diamonds) and D−2 (triangles) endcap disks. The lines represent the calculated values along the lines across the centres of the flux loops.

The errors of the magnetic flux density measured with the flux loops include the standard deviation in the set of seven measurements (7.6 ± 5.0 mT or 0.59% ± 0.32% on average) and a systematic error of ±3.6% arising from the flux loop conductor arrangement. The difference between the modelled and the measured magnetic flux density in the 3D B-sensor locations was 3% ± 7%. The error bars of the 3D B-sensor measurements were ± (0.02 ± 0.01) mT.

## 4. Discussion

After the latest measurements, comparisons of the calculated values of the magnetic flux density in the yoke steel blocks and the measured values obtained in 2006 with the fast discharges of the magnet [33] have been revised. The magnet discharge of November 30, 2017, shown in Figure 13a, was used to exclude the eddy current contributions from the induced voltages of the 2006 measurements. Based on the magnet fast discharge made on November 30, 2017, from the current of 9.5 kA (2 T central magnetic flux density), the eddy current contributions to the 2006 measurements are estimated to be 6.3% ± 4.5% in the barrel wheels and 5.9% ± 3.2% in the endcap disks on average. The revised differences between the calculations done with the latest CMS magnet model [1] and the 2006 measurements are as follows: −0.2% ± 6.8% in the barrel wheels and 0.1% ± 1.6% in the endcap disks at a maximum current of 17.55 kA; 1.4% ± 6.5% in the barrel wheels and 1.3% ± 1.8% in the endcap disks at a maximum current of 19.14 kA. This is compatible with the latest measurements at the operational current of 18.164 kA [30,38].

An attempt to reduce the eddy current contribution with an integration of the voltages induced in the flux loops during the standard magnet ramps has been made before upgrading the flux loop readout system but gave very large errors due to the reading of very small voltages with the previous 12-bit DAQ modules. An upgrade of the readout electronics and a revision of the areas enclosed by the flux loops made it possible to use the standard ramps of the CMS magnet to avoid the large eddy current contribution [30,38]. Stability of these measurements confirmed the correctness of the CMS magnetic field description calculated with the CMS magnet 3D model [1].

## 5. Conclusions

The creation of systems for measuring and monitoring the magnetic field of the CMS detector became possible thanks to the work of many professionals. Each of the systems required a preliminary study of the problem, development work, choice of solutions and ingenuity in the implementation of the entire project.

The success in describing the magnetic flux distribution in the entire CMS detector using a three-dimensional magnet model confirmed by numerous magnetic field measurements with various types of detecting elements was made possible primarily due to the correct concept of dimensions and magnetic induction of the heart of the CMS magnet—a superconducting solenoid. The unique parameters of the solenoid ensured an almost constant and homogeneous magnetic field in the CMS tracking volume. Measurement of the magnetic flux density in the steel blocks of the magnet yoke using flux loops and three-dimensional B-sensors confirmed the correctness of the magnetic flux distribution modelling in the muon momenta measuring system, which provided a high muon momentum resolution and a reliable muon identification.

**Author Contributions:** Technical coordination, A.H., A.B. and W.Z.; conception, V.K.; engineering integration, B.C., D.D., A.G., H.G., R.L. and D.W.; B-sensor electronics, H.B.; B-sensor calibration, F.B.; B-sensor read-out software, F.B. and H.B.; NMR probe measurements, V.K. and B.C.; fieldmapper measurements, V.K., G.T. and J.Z.; flux loop measurements, V.K., B.C., G.T. and J.Z.; B-sensor monitoring system, V.K. and B.C.; data analysis, V.K.; writing—original draft preparation, V.K.; writing—editing, H.B. All authors have read and agreed to the published version of the manuscript.

**Funding:** This research received no external funding.

**Institutional Review Board Statement:** Not applicable.

**Informed Consent Statement:** Not applicable.

**Data Availability Statement:** Not applicable.

**Conflicts of Interest:** The authors declare no conflict of interest.

**References**

1. Klyukhin, V. Design and Description of the CMS Magnetic System Model. *Symmetry* **2021**, *13*, 1052. https://doi.org/10.3390/sym13061052.
2. CMS Collaboration. The CMS experiment at the CERN LHC. *Jinst* **2008**, *3*, S08004. https://doi.org/10.1088/1748-0221/3/08/S08004.
3. Evans, L., Bryant, P. LHC Machine. *J. Instrum.* **2008**, *3*, S08001. https://doi.org/10.1088/1748-0221/3/08/S08001.

4.  Hervé, A. Constructing a 4-Tesla large thin solenoid at the limit of what can be safely operated. *Mod. Phys. Lett. A* **2010**, *25*, 1647–1666. https://doi.org/10.1142/S0217732310033694.
5.  Kircher, F.; Bredy, P.; Calvo, A.; Curé, B.; Campi, D.; Desirelli, A.; Fabbricatore, P.; Farinon, S.; Hervé, A.; Horvath, I.; et al. Final Design of the CMS Solenoid Cold Mass. *IEEE Trans. Appl. Supercond*. **2000**, *10*, 407–410. https://doi.org/10.1109/77.828259.
6.  Herve, A.; Blau, B.; Bredy, P.; Campi, D.; Cannarsa, P.; Curé, B.; Dupont, T.; Fabbricatore, P.; Farinon, S.; Feyzi, F.; et al. Status of the Construction of the CMS Magnet. *IEEE Trans. Appl. Supercond*. **2004**, *14*, 542–547. https://doi.org/10.1109/TASC.2004.829715.
7.  Campi, D.; Curé, B.; Gaddi, A.; Gerwig, H.; Hervé, A.; Klyukhin, V.; Maire, G.; Perinic, G.; Bredy, P.; Fazilleau, P.; et al. Commissioning of the CMS Magnet. *IEEE Trans. Appl. Supercond*. **2007**, *17*, 1185–1190. https://doi.org/10.1109/TASC.2007.897754.
8.  Klyukhin, V.; Ball, A.; Bergsma, F.; Campi, D.; Curé, B.; Gaddi, A.; Gerwig, H.; Hervé, A.; Korienek, J.; Linde, F.; et al. Measurement of the CMS Magnetic Field. *IEEE Trans. Appl. Supercond*. **2008**, *18*, 395–398. https://doi.org/10.1109/TASC.2008.921242.
9.  *TOSCA/OPERA-3d 18R2 Reference Manual*; Cobham CTS Ltd.: Kidlington, UK, 2018; pp. 1–916.
10. Klyukhin, V.I.; Amapane, N.; Andreev, V.; Ball, A.; Curé, B.; Hervé, A.; Gaddi, A.; Gerwig, H.; Karimaki, V.; Loveless, R.; et al. The CMS Magnetic Field Map Performance. *IEEE Trans. Appl. Supercond*. **2010**, *20*, 152–155. https://doi.org/10.1109/TASC.2010.2041200.
11. Klyukhin, V.I.; Amapane, N.; Ball, A.; Curé, B.; Gaddi, A.; Gerwig, H.; Mulders, M.; Hervé, A.; Loveless, R. Measuring the Magnetic Flux Density in the CMS Steel Yoke. *J. Supercond. Nov. Magn*. **2012**, *26*, 1307–1311. https://doi.org/10.1007/s10948-012-1967-5.
12. Klyukhin, V.; Campi, D.; Curé, B.; Gaddi, A.; Gerwig, H.; Grillet, J.P.; Hervé, A.; Loveless, R.; Smith, R.P. Developing the Technique of Measurements of Magnetic Field in the CMS Steel Yoke Elements with Flux-loops and Hall Probes. *IEEE Trans. Nucl. Sci*. **2004**, *51*, 2187–2192. https://doi.org/10.1109/TNS.2004.834722.
13. *NMR Teslameter PT 2025—User's Manual*; METROLAB Instruments SA, GMW Associates: San Carlos, CA, USA, 2003. Available online: https://gmw.com/wp-content/uploads/2019/03/ML_MAN_PT2025-V2-R1-Sep-03.pdf (accessed on 13 January 2022).
14. Sanfilippo, S. Hall probes: physics and application to magnetometry. In Proceedings of the CAS–CERN Accelerator School: Magnets, Bruges, Belgium, 16–25 June 2009; Brandt, D.; arXiv:1105.5069; CERN-2010-004; CERN: Geneva, Switzerland, 2010; pp. 439, 448–449. ISBN 978–92–9083–355-0. Available online: http://cdsweb.cern.ch/record/1158462/files/cern-2010-004.pdf (accessed on 13 January 2022).
15. Boterenbrood, H. *B-Sensor with Addressable Serial Peripheral Interface: User Manual*; Version 1.5; NIKHEF: Amsterdam, The Netherlands, 2001. Available online: http://www.nikhef.nl/pub/departments/ct/po/html/Bsensor/Bsensor.pdf (accessed on 13 January 2022).
16. Bergsma, F.; Blanc, P.H.; Garnier, F.; Giudici, P.A. A High Precision 3D Magnetic Field Scanner for Small to Medium Size Magnets. *IEEE Trans. Appl. Superconduct*. **2016**, *26*, 9000204. https://doi.org/10.1109/TASC.2016.2520204.
17. Bergsma, F. Calibration of Hall sensors in three dimensions. Presented at the 13th Int. Magnetic Measurement Workshop, May 19–22, 2003, Stanford, CA, USA; SLAC: Stanford, CA, USA, 2003. Available online: https://www.slac.stanford.edu/cgi-bin/getdoc/slac-wp-029-ch13-Bergsma.pdf (accessed on 13 January 2022).
18. Bergsma, F. Progress on the 3D calibration of hall probes. Presented at the 14th Int. Magnetic Measurement Workshop 26–29 September 2005, Geneva, Switzerland; CERN: Geneva, Switzerland, 2005. Available online: https://indico.cern.ch/event/408147/contributions/980914/ (accessed on 13 January 2022).
19. Bergsma, F. 3D Hall probes applications. Presented at the15th Int. Magnetic Measurement Workshop, August 21–24, 2007, Batavia, IL, USA; FNAL: Batavia, IL, USA, 2007. Available online: http://fxb.web.cern.ch/fxb/immw15/3d_HallProbes_Applications_Bergsma.pdf (accessed on 13 January 2022).
20. *LabVIEW Software*; National Instruments: Austin, TX, USA, 2005.
21. Boterenbrood, H. *CANopen: High-Level Protocol for CAN-Bus*; Version 3.0; NIKHEF: Amsterdam, The Netherlands, 2000. Available online: http://www.nikhef.nl/pub/departments/ct/po/doc/CANopen30.pdf (accessed on 13 January 2022).
22. Boterenbrood, H. *MDT-DCS CANopen Module: User Manual & reference*; Version 2.6; NIKHEF: Amsterdam, The Netherlands, 2008. Available online: https://www.nikhef.nl/pub/departments/ct/po/html/MDT/MDT-DCS-CANnode-v26.pdf (accessed on 13 January 2022).
23. SIEMENS Semiconductor Group. *Hall Sensor KSY44 Datasheet*; Siemens Corporation: Munich, Germany, 1998. Available online: http://www.hoeben.com/productdata/Siemens%20KSY44%20Datasheet.pdf (accessed on 13 January 2022).
24. Schubiger, M.; Schindler, H.; Blanc, F.; Lindner, R.; Martinelli, M.; Sainvitu, P. Magnetic Feld Map with 2014 Measurements for the LHCb Dipole Magnet. LPHE Note 2015-03, Lausanne: LPHE, 2016. Available online: https://lphe.epfl.ch/publications/2015/LPHE-2015-003_v2.pdf (accessed on 13 January 2022).
25. Tikhonov, A.N.; Samarsky, A.A. *Equations of Mathematical Physics*, 4th ed.; Nauka: Moscow, Russia, 1972; pp. 671–694. (In Russian)
26. Tchébychew, P. Théorie des mécanismes connus sous le nom de parallélogrammes (lu dans l'assemblée le 28 janvier 1853). *Mém. des sav. étrang. prés. à l'Acad. de St. Pétersbourg* **1854**, *7*, 537–568. Available online: https://www.biodiversitylibrary.org/page/37115497#page/579/mode/1up (accessed on 13 January 2022).
27. Klioukhine, V.; Smith, R.P.; Curé, B.; Lesmond, C. *On a Possibility to Measure the Magnetic Field Inside the CMS Yoke Elements*; CMS NOTE 2000/071; CERN: Geneva, Switzerland, 2000. Available online: http://cmsdoc.cern.ch/documents/00/note00_071.pdf (accessed on 13 January 2022).

28. *ELEKTRA/OPERA-3d User Guide*; Version 9.0; Vector Fields Limited: Kidlington, Oxford, UK, 2003, pp. 6-19–6-34.
29. Klyukhin, V.I.; Campi, D.; Curé, B.; Gaddi, A.; Gerwig, H.; Grillet, J.P.; Hervé, A.; Loveless, R.; Smith, R.P. Analysis of Eddy Current Distributions in the CMS Magnet Yoke during the Solenoid Discharge. *IEEE Trans. Nucl. Sci*. **2005**, *52*, 741–744. https://doi.org/10.1109/TNS.2005.850932.
30. Klyukhin, V., Curé, B., Amapane, N., Ball, A., Gaddi, A., Gerwig, H., Hervé, A., Loveless, R., Mulders, M. Using the Standard Linear Ramps of the CMS Superconducting Magnet for Measuring the Magnetic Flux Density in the Steel Flux-Return Yoke. *IEEE Trans. Magn*. **2019**, *55*, 8300504. https://doi.org/10.1109/TMAG.2018.2868798.
31. Landau, L.D.; Lifshitz, E.M. *Electrodynamics of Continuous Media*, 2nd ed.; Nauka: Moscow, Russia, 1982; pp. 154–163. (In Russian)
32. Tamm, I.E. *Fundamentals of the Theory of Electricity*, 9th ed.; Nauka: Moscow, Russia, 1976; pp. 285–288. (In Russian).
33. Klyukhin, V.I., Amapane, N., Ball, A., Calvelli, V., Curé, B., Gaddi, A., Gerwig, H., Mulders, M., Hervé, A., Loveless, R. Validation of the CMS Magnetic Field Map. *J. Supercond. Nov. Magn*. **2015**, *28*, 701–704. https://doi.org/10.1007/s10948-014-2809-4.
34. *USB-1208LS Analog and Digital I/O—User's Guide*; Measurement Computing Corporation, Norton, MA, USA, 2005. Available online: https://www.mccdaq.com/pdfs/manuals/USB-1208LS.pdf (accessed on 13 January 2022).
35. *AnywhereUSB—User's Guide*; Digi International Inc., Minnetonka, MN, USA, 2005. Available online: https://www.digi.com/resources/documentation/digidocs/pdfs/90001085.pdf (accessed on 13 January 2022).
36. *3Com OfficeConnect® Dual Speed Switch 5*; 3Com Corporation, Marlborough, MA, USA, 2005. Available online: https://support.hpe.com/hpesc/public/docDisplay?docId=c02581538&docLocale=en_US (accessed on 13 January 2022).
37. Klyukhin, V.I.; Ball, A.; Campi, D.; Curé, B.; Dattola, D.; Gaddi, A.; Gerwig, H.; Hervé, A.; Loveless, R.; Reithler, H.; et al. Measuring the Magnetic Field Inside the CMS Steel Yoke Elements. In Proceedings of the 2008 IEEE Nuclear Science Symposium Conference Record, Dresden, Germany, 19–25 October 2008; pp. 2270–2273. https://doi.org/10.1109/NSSMIC.2008.4774806.
38. Klyukhin, V.I., Amapane, N., Ball, A., Curé, B., Gaddi, A., Gerwig, H., Mulders, M., Hervé, A., Loveless, R. Flux Loop Measurements of the Magnetic Flux Density in the CMS Magnet Yoke. *J. Supercond. Nov. Magn*. **2017**, *30*, 2977–2980. https://doi.org/10.1007/s10948-016-3634-8.